\documentclass[aps,prl,twocolumn, eqsecnum, groupedaddress]{revtex4}
\usepackage{graphicx}
\begin{document}


\title{ A Collective Heavy Fermion State and Superconductivity in Pr$_{1-x}$La$_x$Os$_4$Sb$_{12}$: Specific Heat and Susceptibility Study.}


\author{ C.R. Rotundu, P. Kumar, and B. Andraka }%
\email{andraka@phys.ufl.edu}
\affiliation{ Department of Physics, University of Florida\\
P.O. Box 118440, Gainesville, Florida  32611-8440, USA}


\date{\today}

\begin{abstract}
Low temperature susceptibility and specific heat for single crystals of Pr$_{1-x}$La$_x$Os$_4$Sb$_{12}$ (0 $\leq x$ $\leq 1$) are reported. La-doping leaves CEF energies of Pr essentially unchanged. The average $T_c$ is only weakly affected by the La-substitution and varies approximately linearly between the end-compounds. The second superconducting transition disappears between $x$=0.05 and 0.1. The discontinuity in $C/T$ at $T_c$, on the other hand, is drastically reduced from about 1000 mJ/K$^2$mol for $x$=0 to 200 mJ/K$^2$mol for $x$=0.2. $\Delta C/T_c$ decreases further for alloys corresponding to $x\geq 0.6$ to the value of a conventional superconductor LaOs$_4$Sb$_{12}$.  This behavior implies non-single impurity origin of the heavy fermion state in PrOs$_4$Sb$_{12}$. We argue that critical quadrupolar fluctuations are responsible for this heavy fermion state.
\end{abstract}

\pacs{71.27.+a, 74.70.Tx, 75.40.Cx}

\maketitle

\section{}
\narrowtext

Lately, there has been great interest in the first discovered Pr-based heavy fermion superconductor\cite{Bauer}, PrOs$_4$Sb$_{12}$. The significance of this system stems from several reasons. Firstly, a large discontinuity in $C/T$ at $T_c$ and correspondingly large temperature slope of the upper critical field ($dH_{c2}/dT$) unambiguously prove that Pr can support heavy fermion state(s). Secondly, the paramagnetic heavy fermion state seems to be unconventional; i.e., not of a magnetic Kondo effect origin. It has been postulated that the J=4 multiplet of the Pr$^{3+}$ ion is split by the crystalline electric field (CEF) such that the ground state is a nonmagnetic $\Gamma_3$ doublet. Since the $\Gamma_3$ doublet carries no magnetic dipole moment but a quadrupolar electric moment, a quadrupolar Kondo effect, a mechanism\cite{Cox} never established experimentally to be relevant for heavy fermions, has been proposed as the source of heavy electrons and superconductivity in PrOs$_4$Sb$_{12}$. However, our recent magnetic field investigation\cite{Rotundu} provided support for a complementary CEF configuration, with a singlet ($\Gamma_1$) ground state and the first excited state being a triplet ($\Gamma_5$), at about 8 K. This model is also consistent with inelastic\cite{Maple} and elastic neutron diffraction\cite{Kohgi}, specific heat\cite{Bauer,Aoki} and magnetic susceptibility data. Thus, the formation of heavy electrons requires participation of the excited crystal field levels. The qudrupolar interactions are also expected to be strong in this $\Gamma_1$ - $\Gamma_5$ model. Upon application of the magnetic field, the singlet crosses the lowest level of the split triplet (at about 8-9 T), forming a pseudodoublet possessing a quadrupolar electric moment. Antiferroquadrupolar order is observed in PrOs$_4$Sb$_{12}$ in magnetic fields\cite{Kohgi,Aoki}, between about 4.5 and 14 T. The strength of quadrupolar interactions peaks at the crossing field. Thus, the $\Gamma_1$ - $\Gamma_5$ CEF model does not apriori preclude the possibility of the superconductivity mediated by quadrupolar interactions.

One of the strongest arguments for the unconventional (anisotropic) superconductivity is provided by the observation of two superconducting transitions in some PrOs$_4$Sb$_{12}$ samples\cite{Vollmer}. And yet, the results of the $\mu SR$ experiment\cite{MacLaughlin} performed on samples with a single $T_c$ suggest isotropic superconducting order parameter. To shed further light on the character of the superconductivity, we have performed a systematic study of the specific heat on the La-doped samples. Since La has no f-electrons, substituting La for Pr should have a strong effect on the superconductivity in the quadrupolar scenario.

All single-crystalline samples used in this investigation were by the Sb-flux method\cite{Bauer2}. La and Pr were premelted several times in an arc-melter to improve the homogeneity of samples. The results of  the X-ray diffraction analysis were consistent with single phase materials. We have detected  a monotonic, but very small, increase of the lattice constant with the La-content. These very small changes (0.03 \% between the end compounds; on the border of sensitivity of our technique) are in an agreement with previously reported\cite{Braun} an almost non-existent lanthanide contraction in ternary skutterudites containing Sb, of a general form LnT$_4$Sb$_{12}$, where T and Ln are transition element and light lanthanide, respectively. For most of the concentrations studied, we have performed measurements, both specific heat and susceptibility, on two different, randomely selected crystals. Within the uncertainty of these measurements, we have observed very good reproducibility for crystals with the same nominal concentration.

\begin{figure}[btp]
\begin{center}
\leavevmode 
\includegraphics[width=0.8\linewidth]{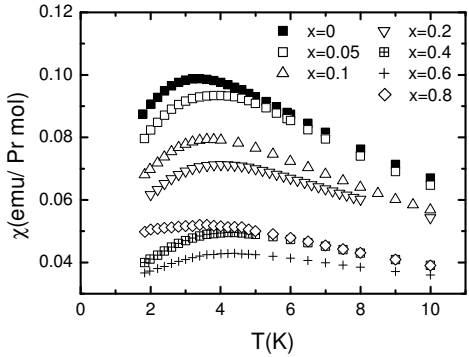} 
\caption{ Magnetic susceptibility of Pr$_{1-x}$La$_x$Os$_4$Sb$_{12}$ between 1.8 and 10 K, measured in the field of 0.5 T.}
\label{fig1}
\end{center}
\end{figure}
Figure 1 shows the susceptibility for Pr$_{1-x}$La$_x$Os$_4$Sb$_{12}$ samples at temperatures 1.85 to 10 K obtained in the field of 0.5 T and normalized to a mole of Pr. All curves show a low temperature maximum that we believe is due to excitations between the singlet and triplet CEF states. This maximum is at approximately 3.4 K and shifts only slightly to higher temperatures with x, such that it is around 4.3 K for x=0.6. Thus, these results indicate that the crystalline electric field scheme of Pr remains unchanged and that average separation between the lowest CEF levels increases only slightly with the La-doping. The maximum value of the susceptibility, however, does not stay constant across the system. Initially, it decreases continuously from about 100 memu/Pr mol for the pure material to about 50 memu/Pr mol for x=0.4 followed by smaller changes for higher concentrations of La. Note that because of the small size of the crystals (mass 1 to 4 mg), the sample signal was comparable to the background (sample holder) below 10 K and was much smaller than the background at room temperature. Therefore, we do not attempt to analyze the data above 10 K. Also, the discrepancy in the 1.8 K susceptibility for the three highest La-content compositions is within the absolute error bar. We observe a broadening of the maximum, particularly for x=0.6 and 0.8 compositions. Such a broadening in mixed alloys is expected because of the La/Pr disorder leading to some distribution of Pr-ligand ion distances and, consequently, to smearing out of sharp CEF levels in the undoped material. Susceptibility curves corresponding to different x's have a clear tendency to converge at some higher temperatures. However, the very large initial drop in the lowest temperature susceptibility (1.8 K) is difficult to reconcile with the disorder only. Quite possibly, some characteristic electronic energy (inversely proportional to temperature, analogous to a Kondo temperature) increases sharply upon substituting La for Pr.
\begin{figure}[btp]
\begin{center}
\leavevmode 
\includegraphics[width=0.8\linewidth]{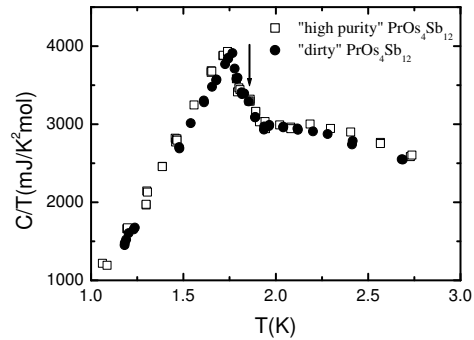} 
\caption{ $C/T$ versus $T$ for PrOs$_4$Sb$_{12}$ samples from two different batches. The sample labeled "dirty" was made of recycled Os. The arrow indicates the second superconducting anomaly.}
\label{fig2}
\end{center}
\end{figure}

The specific heat shown in Figs. 2 and 3 was measured on the same crystals as the susceptibility presented in Fig. 1. From the total specific that corresponding to LaOs$_4$Sb$_{12}$ was subtracted and the result divided by $(1-x)$. LaOs$_4$Sb$_{12}$ has been approximated by the following expression proposed by Bauer et al.\cite{Bauer}, 
 $C$ = 36 $T$ + 1.18 $T^3$; 
where $C$ is in mJ/K mol. Note that the cubic coefficient is significantly smaller than the phonon term derived by Vollmer et al.\cite{Vollmer} for PrOs$_4$Sb$_{12}$. Also, the linear term is smaller than the value reported by Sugawara, et al\cite{Sugawara}. These discrepancies are of some significance when analyzing the data for the largest x-values, particularly for $x=0.8$. Fig. 2 shows such normalized f-electron specific heat divided by temperature ($C/T$) for two crystals of PrOs$_4$Sb$_{12}$ from two different batches obtained in the same manner but with somewhat different purity of starting Os. The crystal labeled "dirty" was grown from recycled Os, which had not been analyzed for impurities. Within the experimental resolution, $C/T$ results are identical. Also, these results are consistent with two superconducting transitions reported by Vollmer et al. The positions of the smaller, higher temperature, peaks are marked with an arrow in Fig. 2. Some smearing out of this higher temperature peak could be due to the measurement method itself, which integrates the specific heat at any temperature $T$ over 0.03 $T$ interval. Two superconducting transitions can still be resolved for $x=0.05$ but not for higher concentrations of La (Fig. 3). Thus, for consistency of the analysis, we treat this double structures in $x=0$ and 0.05 as a single transition. The average $T_c$ for the pure material, approximated using an equal area construction method, is 1.83 K, in a good agreement with the result of Bauer et al.\cite{Bauer} that used a similar procedure. On the other hand, the equal area construction is not reliable for the extraction of $\Delta C/T$ at $T_c$ in the case of a double transition Therefore, we estimate this discontinuity as the measured difference between the maximum value of $C/T$ and the value of $C/T$ just above the transition. This $\Delta C/T$ for $x=0$ is about 1000 mJ/K$^2$mol and among the highest reported, confirming the heavy fermion state. ($\Delta C/T$ associated with the lower temperature, more pronounced transition, is at least 500 mJ/K$^2$mol.) The specific heat near its local maximum around 3 K is about 6.9 J/K mol. This value is in an excellent agreement with that reported by Aoki et al.\cite{Aoki} Note further that this value is significantly larger than that expected for the Schottky maximum corresponding to excitations between the doublet and triplet (about 5.1 J/K mol) and it is smaller than that for the excitations between the singlet and triplet states (8.5 J/Kmol)\cite{Gopal}. However, we do expect some hybridization between the f-electrons of Pr and ligand states leading to the reduction of ionic properties of Pr. Extrapolated values of $T_c$,  $\Delta C/T$, and $C/T$ at its maximum (2.2 - 2.5 K) are also shown in Table 1.

Figure 3 displays the f-electron contribution to $C/T$ versus $T$ for the remaining Pr$_{1-x}$La$_x$Os$_4$Sb$_{12}$ alloys. There are several important contributions to the uncertainty of the data, such as the aforementioned phonons, normal electrons, and the addenda that become more critical for larger values of x. Therefore, we estimate this uncertainty in the f-electron contribution to $C/T$ to increase from about 10 \% for $x=0$ to 25 \% for $x=0.8$. Note further that our procedure of calculating the f-electrons specific heat by subtracting the normal state specific heat of LaOs$_4$Sb$_{12}$ is incorrect for the superconducting state and can lead to unphysical negative values at low temperatures. The Schottky maximum, which is seen near 2.2 K in $C/T$, is indeed smeared out for $x>0$ and moved to slightly higher temperatures, as expected from the susceptibility data. The $C/T$ value at this shallow maximum is reduced by a factor of 2 between the pure compound and alloys corresponding to $x=0.6$ and 0.4 (see also Table 1). Thus, the size of the anomaly in $C/T$(at 2.2 - 2.5 K) scales roughly with the corresponding low temperature maximum in the low temperature susceptibility.
 
\begin{figure}[btp]
\begin{center}
\leavevmode 
\includegraphics[width=0.8\linewidth]{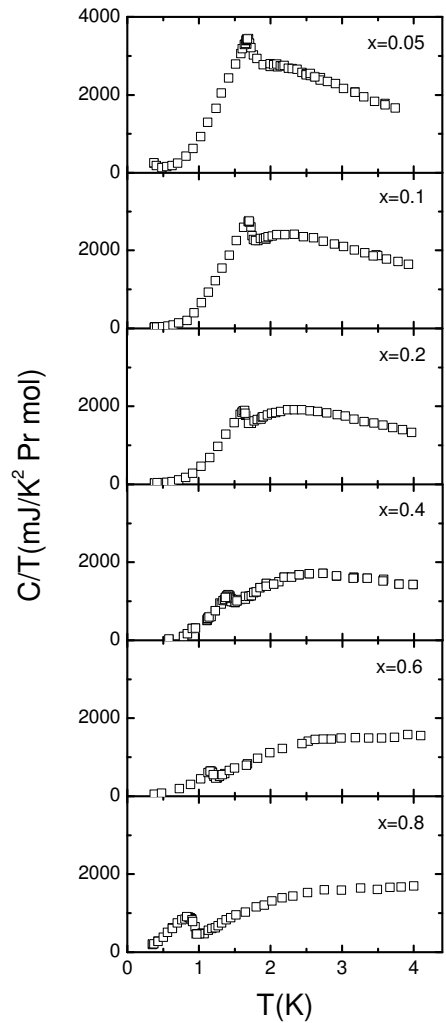} 
\caption{ $C/T$ versus $T$ for Pr$_{1-x}$La$_x$Os$_4$Sb$_{12}$, $x>0$. }
\label{fig3}
\end{center}
\end{figure}
As it can be inferred from Fig. 3, the superconducting transition is only moderately suppressed by the La substitution. This is more clearly shown in Fig 4 and Table 1. We include the published result\cite{Bauer2, Sugawara} for LaOs$_4$Sb$_{12}$, material that is also superconducting below about 0.74 K. $T_c$ varies approximately linearly between the end-compounds. Thus, these results seem to imply that La impurities are not strong pair-brakers and there is a smooth evolution between the superconducting states of PrOs$_4$Sb$_{12}$ and LaOs$_4$Sb$_{12}$. Interestingly, the width of the superconducting transition in $C/T$, measured in a consistent manner for all samples, is the same for $x=0$ and 0.05 and is reduced by a factor of 3 for all the remaining alloys. This is quite an unexpected result; the transition is sharp for all concentrations and becomes even narrowor for mixed alloys. The sharpness of the transition for all concentrations could be due to very small variation of the lattice constant. The reduction of the width is because of the disapearance of one of the superconducting anomalies. Although additional studies are needed on samples with very small amount of La, a closer inspection of superconductive anomalies for $x=0$, 0.02 (not shown), and 0.05 suggests that the La-doping suppresses mainly the lower temperature anomaly in $C/T$.

This approximately linear variation of the average $T_c$ on $x$ in Pr$_{1-x}$La$_x$Os$_4$Sb$_{12}$ is unusual for heavy fermion alloys. For instance, UBe$_{13}$ that shows a number of striking similarities to PrOs$_4$Sb$_{12}$ has its superconductivity suppressed by just 3 \% of La\cite{Ahlheim}. This very different sensitivity of $T_c$ on La-impurities between PrOs$_4$Sb$_{12}$ and canonical heavy fermion superconductors cannot be accounted for by vastly different coherence lengths. In fact, the coherence lengths of the UBe$_{13}$ and PrOs$_4$Sb$_{12}$ are almost identical, 140 (ref.\cite{Maple2}) and 120 $\r{A}$ (ref.\cite{Bauer}), respectively. A somewhat stronger supression of $T_c$ in PrOs$_4$Sb$_{12}$ was found for Ru impurities replacing Os\cite{Fredrick}. In addition, Pr(Os$_{1-x}$Ru$_x$)$_4$Sb$_{12}$ shows a shallow minimum in $T_c$ near $x=0.6$. However, even in this case the rate of the reduction of $T_c$ is small in comparisson with majority of Ce- and U-based heavy fermions  and considering very different paramagnetic states of PrOs$_4$Sb$_{12}$ and PrRu$_4$Sb$_{12}$. PrRu$_4$Sb$_{12}$ is quite an ordinary metal\cite{Takeda} with a different CEF scheme of Pr and relevant CEF energies much larger than those for PrOs$_4$Sb$_{12}$.

Despite the weak dependence of $T_c$ in PrOs$_4$Sb$_{12}$ on La impurities, the discontinuity in $C/T$ (and $C$) at $T_c$ is strongly reduced with $x$ (Table 1). An apparent increase of  $\Delta C/T_c$ between $x=0.6$ and 0.8 in Fig. 3 is because of the normalization by the Pr concentration used in this figure. Such a normalization becomes incorrect for sufficiently large $x$ -values, since LaOs$_4$Sb$_{12}$ itself is superconducting and thus displays this discontinuity. Fig. 4 shows both $T_c$ and unnormalized $\Delta C/T_c$ versus $x$. $\Delta C/T_c$ decreases from approximately 1000 mJ/K$^2$mol ($x=0$) to about 460, and 210 mJ/K$^2$mol when only 10, and 20 \% of La is substituted for Pr, respectively. Lack of normalization to a mole of Pr can not account for this dramatic reduction. Alloys corresponding to $x\geq 0.6$ have $\Delta C/T_c$ approximately equal to that of a conventional superconductor LaOs$_4$Sb$_{12}$. In BCS-type superconductors, $\Delta C/T_c$ is simply related to the electronic specific heat coefficient $\gamma$. There is no such a direct relationship in heavy fermion metals exhibiting multiple superconducting transitions. In some thoriated UBe$_{13}$ samples\cite{Ott}, showing two superconducting transitions, $\Delta C/T_c$ is enhanced by a factor of 2 with respect to UBe$_{13}$. This enhancement seems to be due to some collective excitations present in the normal state and condensing at $T_c$. Thus, similar fluctuations can be responsible for the enhanced $\Delta C/T_c$ in the pure PrOs$_4$Sb$_{12}$ over the alloys with a single transition. On the other hand, this strong variation of $\Delta C/T_c$ on $x$ among alloys with a single transition ($x\geq 0.1$) can be associted with the suppression of $\gamma$ upon dilution with La. Therefore, this behavior clearly implies non-single impurity origin of the heavy fermion state in PrOs$_4$Sb$_{12}$. The electronic specific heat coefficient is reduced to that of LaOs$_4$Sb$_{12}$ by diluting Pr with La. 
 
\begin{figure}[btp]
\begin{center}
\leavevmode 
\includegraphics[width=0.8\linewidth]{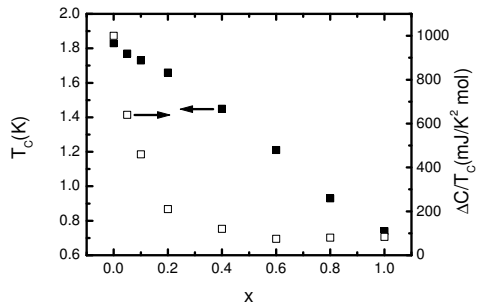} 
\caption{ Superconducting transition temperature $T_c$ (left scale) and $\Delta C/T_c$ (right scale) versus concentration $x$ for Pr$_{1-x}$La$_x$Os$_4$Sb$_{12}$, where 0 $\leq x$ $\leq 1$. $\Delta C$ is a discontinuity in the specific heat at $T_c$.}
\label{fig4}
\end{center}
\end{figure}
Proving or disproving single-ion behavior using just one kind of alloying is, in general, a difficult tusk because alloying can alter parameters of a single-ion Hamiltonian as well. However, in the Pr$_{1-x}$La$_x$Os$_4$Sb$_{12}$ case, the relevant single-ion parameters seem to be unaffected or only slightly affected by alloying, probably due to the peculiarity of the crystal structure. The lattice constant, CEF scheme of Pr, and the lowest CEF energy show very weak sensitivity to the La-doping. On the other hand, the alloying is expected to reduce qudrupolar fluctuations shown to be strong in PrOs$_4$Sb$_{12}$. As it has been discussed, magnetic fields 4.5 T and larger induce AFQ state. Such a state is known to be very susceptible to alloying. For instance, in PrPb$_3$, undergoing AFQ ordering in zero field at 0.4 K, just 2 \% of La substituted for Pr supresses the long range order completely\cite{Kawae}. Our, preliminary investigation of Pr$_{0.8}$La$_{0.2}$Os$_4$Sb$_{12}$ in magnetic fields, has not found any evidence of a field-induced AFQ order above 0.4 K. Thus, the heavy fermion state in PrOs$_4$Sb$_{12}$ seems to correlate with the field-induced AFQ order, or proximity to AFQ order.

\begin{table}
\begin{center}
\vspace{1ex}
\caption{Specific heat parameters for Pr$_{1-x}$La$_x$Os$_4$Sb$_{12}$ alloys, 
         $0\leq\,x\leq\,1$.}
\label{table1}
\begin{tabular}{cccc}
 La $x$  & $T_c$ (K) & $\Delta C/T_c$ (mJ/K$^2$mol) & $C/T$ max (mJ/K$^2$Pr mol) \\

 0  & 1.83 & 1000 & 2900 \\
 0.05 & 1.77 & 640 & 2700 \\
 0.1 & 1.73 & 460 & 2400 \\
 0.2 & 1.66 & 210 & 1600 \\
 0.4 & 1.45 & 120 & 1700 \\
 0.6 & 1.20 & 70 & 1500 \\
 0.8 & 0.93 & 80 & 1900 \\
 1 & 0.74\cite{Bauer2,Sugawara} & 84\cite{Sugawara} & - \\
 \end{tabular}
\end{center}
\end{table}
Interestingly, there is no correlation between the heavy fermion character measured by $\Delta C/T_c$ and the average $T_c$. Our results argue against the same mechanism responsible for the heavy fermion state and enhanced value of $T_c$ in PrOs$_4$Sb$_{12}$. Changes in the phonon spectrum are most probably behind the variation of $T_c$ in Pr$_{1-x}$La$_x$Os$_4$Sb$_{12}$. In fact, Vollmer et al. derive a much lower Debye temperature for PrOs$_4$Sb$_{12}$ than that for LaOs$_4$Sb$_{12}$, from the specific heat data at tempertures to 10 K. Although the average $T_c$ seems to be uncorrelated with the Sommerfeld coefficient, the superconducting state in PrOs$_4$Sb$_{12}$ is clearly affected by the heavy fermion state. In particular, a second superconducting transition and spontaneous magnetic field show up\cite{Aoki2} below $T_c$ in Pr$_{1-x}$La$_x$Os$_4$Sb$_{12}$ alloys with a heavy fermion normal state. This superconductivity provides also a proof of the heavy fermion state in PrOs$_4$Sb$_{12}$.  
\begin{acknowledgments}
This work has been supported by the U.S. Department of Energy, Grant No. DE-FG02-99ER45748, National Science Foundation, DMR-0104240. We thank G.R. Stewart and Y. Takano for stimulating discussions.
\end{acknowledgments}


\end{document}